\documentclass[prd,psfig,twocolumn]{revtex4}
\usepackage{psfig}
\usepackage{bm}
\usepackage{amsmath,amssymb,amsfonts}
\def\be{\begin{eqnarray}}
\def\ee{\end{eqnarray}}
        
\newcommand{\vecv}{\bm v}        
\newcommand{\vecn}{\bm n}        
\newcommand{\vecu}{\bm u}        
\newcommand{\vecf}{\bm f}        
\newcommand{\vecp}{\bm p}        
\newcommand{\vecB}{\bm B}

\def\bnabla{{\mbox{\bm $\nabla$}}}

\begin{document}        
\title{
Type-I superconductivity and neutron star precession}
\author{Armen Sedrakian}
\affiliation{Institute for Theoretical Physics, T\"ubingen 
University, 72076 T\"ubingen, Germany
}
\begin{abstract}
Type-I proton superconducting cores of neutron stars  break up in a magnetic
field into alternating domains of superconducting and normal fluids. 
We examine two channels of superfluid-normal fluid friction where 
(i) rotational vortices are decoupled from the non-superconducting 
domains and the interaction is due to the strong force between protons 
and neutrons; (ii) the non-superconducting domains are dynamically
coupled to the vortices and the vortex motion generates transverse 
electric fields within them, causing electronic current flow and Ohmic 
dissipation. The obtained dissipation coefficients are consistent with 
the Eulerian precession of neutron stars.
\end{abstract}
\date{\today}

\maketitle

\section{Introduction}
The timing observations of radio-pulsars provide a unique tool to 
study the properties of superdense matter in compact stars. 
While pulsars are known to be prefect clocks over long periods 
of time, the timing observations of past few decades
revealed several types of timing ``irregularities'' in a 
subclass of isolated compact objects. 
Glitches (or macrojumps) - sudden increases
in the pulsar rotation frequency and its derivative - are the most
spectacular examples of timing anomalies. The slow relaxation of their 
spin and its derivative following a glitch has been interpreted
as an evidence for superfluidity of compact star interiors~\cite{CARTER}.
At temperatures prevailing in an evolved compact star the dense 
hadronic matter is expected to be in the superfluid state due
to the attractive component of the nuclear force which binds 
neutrons into Cooper pairs 
either in the relative $^1S_0$ state (at low densities) or 
$^3P_2-^3F_2$ state (at high densities) \cite{MORTEN}. Other members
of the spin-1/2 octet of  baryons (protons 
and strangeness $S =1$ particles $\Sigma^{-0+}$, $\Lambda$) 
are expected to pair in the $S$-wave channel due to their 
low concentration. If the central densities 
of compact stars exceed the density of deconfinement phase transition
to a quark matter phase, the deconfined quark matter will be in 
one of the many possible color superconducting states \cite{ALFORD}. 
Since the glitches are related to the axisymmetric perturbations
from the state of uniform rotation there is a twofold degeneracy 
in the interpretation of the data on both the jumps and the 
post-jump relaxations; the time-scales of these processes can  
be associated either with the weak or strong coupling between 
the superfluid and normal fluid in the star's interiors, and it 
is impossible to distinguish between these regimes on the basis
of glitch observations alone. During the recent years it became 
increasingly clear that another type of timing anomaly - the 
long term periodic variations superimposed on the spin-down of the
star -  can provide an additional and independent information on the dynamical
coupling between the superfluid and the normal fluid in compact stars.
If these irregularities are interpreted in terms of the precession 
of the star (a motion which involves non-axisymmetric perturbations
from the rotational state) the degeneracy
inherent to the interpretation of glitches is lifted. It turns 
out that the free precession 
is possible only in the weak coupling limit and it is damped 
in the strong coupling case. Interpretations of the timing anomalies
in pulsars in terms of the friction between the superfluid and
the normal fluid 
require a model of the friction between the superfluid 
and normal components of the star on mesoscopic scales characteristic 
for the vorticity. This paper discusses two new mechanism of  
mutual friction in the core of a neutron star in the 
case where protons form a type-I superconductor. The remainder 
of the introduction sets the stage by briefly reviewing the relevant 
physics. Section \ref{1+1} studies the dynamics 
of a  type-I superconducting model where there is a single 
normal domain per rotational vortex. In section \ref{1+N} we discuss the 
dissipation in an alternative picture where there is a large number of 
rotational vortices associated with a single normal domain.
Section \ref{conclude} is devoted to the implications of the
dissipative dynamics of type-I superconductors for the free precession 
of compact stars and contains a brief summary of the results.

\subsection{No-go theorems for precession}

To see how the superfluidity of neutron stars changes
their Eulerian precession (which would be intact if 
the neutron stars were non-superfluid) let us begin with the
equation of motion of approximately massless neutron vortex  
\be\label{eq:1} 
\rho_S\kappa\,\, (\vecv_S-\vecv_L)\times \vecn 
-\eta\vecu-\eta'(\vecn\times\vecu) = 0,
\ee
where $\vecu\equiv\vecv_N-\vecv_L$, and $\vecv_S$, $\vecv_N$ 
and $\vecv_L$ are  
the velocities of the superfluid, the normal fluid  and  the vortex;
 $\rho_S$ is the effective neutron density, $\kappa$ is the 
quantum of circulation, $\vecn= \vec \kappa /\kappa$,
and the coefficients $\eta$ and $\eta'$ are the measure of the
friction between the neutron vortex and the ambient normal fluid. 
Here we work 
within the two-fluid superfluid hydrodynamics, where it is assumed
that the hydrodynamic forces are linear functions of the velocities,
which guarantees that the energy variation is always a quadratic
form. Shaham first observed that the long-term,
Eulerian precession is impossible if the neutron vortices are 
strongly pinned \cite{SHAHAM}. In terms of the mesoscopic parameters 
in Eq. (\ref{eq:1}), his observation is equivalent to the statement  that 
precession is absent in the limit $\zeta\to \infty$,  $\zeta'\to 0$ where
$\zeta = \eta / \rho_S\kappa$  and $\zeta' = \eta' / \rho_S\kappa$
are the drag-to-lift ratios.
Since in the frictionless limit a star must precess at 
the classical frequency $\epsilon\Omega$, where $\epsilon$ is the eccentricity
and $\Omega$ is the rotation frequency, it is clear that there exists 
a crossover from the damped to the free precession as $\zeta$ is decreased.
The crossover is determined by the dimensionless parameters
($I_S/I_N)\,\beta$  and ($I_S/I_N)\,\beta'$  
where $I_S$ is the moment of inertia of the superfluid 
and $I_N$ is the moment of inertia of the crust plus any 
component coupled to it on time-scales much shorter than 
the precession timescale and $\beta = \zeta/[(1-\zeta')^2+\zeta^2]$, 
 $\beta' = 1- \beta(1-\zeta')/\zeta$. The precession 
frequency is~\cite{SWC} (hereafter SWC) 
\be\label{PRECESSION}
\Omega_P = \epsilon\Omega_S\left[ \left(1+\beta'\frac{I_S}{I_N}\right)
+i \beta \frac{I_S}{I_N}\right],
\ee
where $\Omega_S$ is the spin frequency and $\epsilon$ is the eccentricity.
The result of SWC can be cast in 
a no-go theorem that states that the Eulerian precession in a superfluid 
neutron star is impossible if ($I_S/I_N)\,\zeta >1$ (assuming as
before $\zeta'\to 0$). 
There is a subtlety to this result:  the precession 
is impossible because the precession mode, apart from being damped,
is renormalized by the non-dissipative component of superfluid-normal 
fluid interaction ($\propto \beta')$. In effect the value of the precession 
eigen-frequency drops  below the damping frequency for any $\zeta$
larger than the crossover value.  Note that 
this counter-intuitive result can not be obtained from the
arguments based solely on dissipation: in fact, according to 
Eq. (\ref{PRECESSION}) the damping time-scale for 
precession increases linearly with $\zeta$ and in the limit 
$\zeta\to \infty$ one would  predict (wrongly) undamped precession.
If a neutron star contains multiple layers
of superfluids the picture is more complex, but the generic features 
of the crossover are the same \cite{SWC}.

\subsection{Previous work}

Long term variabilities were observed in a number of pulsars
and have been attributed phenomenologically to precession of the
neutron star (see ref. \cite{CORDES} and references therein). 
A strong case for long-term variability (again attributed to 
precession) was made recently by Stairs et al \cite{STAIRS}. While it
is common to study  perturbations from the state of uniform
rotation,  Wasserman \cite{WASSERMAN} demonstrated that the
precessional state may correspond to the local energy minimum 
of an inclined rotator if there is a large enough magnetic stress on the
star's core. This type of precession is likely to be damped away by the 
superfluid-normal friction.
Link  \cite{BLINK} argued that the long-term variations, which can be fitted
by assuming Eulerian precession of the pulsar, 
are incompatible with type-II superconductivity of 
neutron stars. 
Type-I superconductivity was proposed to resolve the discrepancy
\cite{BLINK}. Jones \cite{JONES} argued that the friction of vortices
in the crusts against the nuclear lattice will give rise to a
dissipation which will damp the free precession; thus, the free
precession (even in absence of pinning) would be incompatible with 
the known properties of matter at subnuclear densities.

\subsection{Type-I superconducting neutron stars}

As is well known, type-I superconductivity arises when the Ginzburg-Landau
parameter satisfies the condition $\kappa_{GL}= \lambda/\xi \le 1/\sqrt{2}$,
where $\lambda$ is the magnetic field penetration depth and 
$\xi$ is the coherence length.
Type-I superconductivity can arise locally within the current models based 
on the BCS theory \cite{SSZ}, with domain structures analogous to 
those observed in  laboratory experiments. In ref. \cite{SSZ}
the theory of these structures was constructed along the lines 
of the theories developed for laboratory superconductors, 
where the magnetic fields are  generated 
by normal currents driven around a cylindrical cavity by temperature 
gradients \cite{GZ}. However, global type-I superconductivity would 
require a  suppression of the proton pairing gap $\Delta_p$  
(due to the scaling $\kappa_{GL}\propto \xi^{-1}
\propto \Delta_p$) 
by polarization or related effects. 
Buckley et al \cite{BUCKLEY} studied the effect the 
interactions between the neutron and proton Cooper pairs 
would have on the type of the proton superconductivity.
Their results suggest that type-I superconductivity can be 
enforced within the entire core
without the suppression of the pairing gap if 
the strength of the yet unknown interaction between Cooper 
pairs will turn out to be significant.

The equilibrium structure of the alternating superconducting
and normal domains in a type-I superconductor is a complicated problem
and depends, among other things, 
on the nucleation history of the superfluid phase. 
The equilibrium dimension of a layer is of the order of magnitude 
$d\sim \sqrt{L\xi}$ where $L$ is the size of the core; (for typical parameter
values $L\simeq 5\times 10^5$ cm and $\xi \simeq 200$ fm, $d\sim 3.2
\times 10^{-3}$ cm). By flux conservation, 
the ratio of the sizes of the superfluid and normal 
domains is given by the relation $d_S/d_N = \sqrt{H_{\rm cm}/B}\sim 10$, 
where $B\sim 10^{12}$ G is the average value of the magnetic induction, 
$H_{\rm cm}\sim 10^{14}$ G
is the thermodynamic magnetic field. The dimensional analysis 
above suggests that there is roughly a single normal domain per neutron
vortex. We shall consider below the dynamics of two distinct models where
(i)  a neutron vortex features a single coaxial normal domain of a 
smaller size according to ref.  \cite{SSZ} (hereafter $1+1$ model) 
and (ii) a large fraction of neutron 
vortices is accommodated by a single normal domain, as described in 
ref. \cite{BUCKLEY} ($1+N$ model). The difference between 
these models is not simply the number of the neutron vortices
accommodated by a single non-superconducting domain; 
in the 1+1 model the magnetic fields
are generated dynamically (this is explained in more detail below) and hence
the normal domains are tied to the neutron vortices on dynamical times
scales. In effect the motion of a neutron vortex requires the motion 
of the normal domain attached to it on the dynamical timescale. Therefore
the dissipation is due to the interaction between the combined structure 
(a vortex plus a normal domain) with the background electron liquid.
While it is conceivable that the normal domains are tied up to the 
macroscopic neutron vortex lattice on certain timescales in 
the $1+N$ model, we shall assume that the neutron vortex lattice sweeps 
through the non-superconducting domains and the dissipation arises from 
the scattering of the normal protons off the cores of rotational
vortices \cite{SEDRAKIAN98}.

\section{Dynamics of 1+1 structures}
\label{1+1}

In two component superfluids the supercurrent of any given 
component transports  mass of both components; this is the essence of 
the entrainment effect first studied in the context of charge neutral 
and non-rotating superfluid mixtures of $^3$He-$^4$He \cite{ANDREEV}.
If one of the components is charged - a case first studied in 
refs. \cite{SEDRAKIAN,ALPAR} -  the neutral superfluid which rotates
by forming a lattice of charge-neutral vortices generates magnetic 
fields because neutral supercurrent carries along a finite mass 
of the charged component. The mass currents of neutrons 
and protons $\vecp_{p/n}$ (here and below 
indices $p$ and $n$ refer to the protons and neutrons), 
are related to their velocities $\vecv_{n}$ and $\vecv_{p}$ by 
a density matrix in the isospin space \cite{ANDREEV,SEDRAKIAN,ALPAR}
\be 
\left(
\begin{array}{c}
\vecp_p\\
\vecp_n\\ 
\end{array}\right) = \left(
\begin{array}{cc}
\rho_{pp}&\rho_{pn}\\
\rho_{np}&\rho_{nn}\\
\end{array}\right)
\left(\begin{array}{c}
\vecv_p\\
\vecv_n\\ 
\end{array}\right),
\ee 
where in the mean-field approximation the elements of the density 
matrix can be expressed through the effective masses of neutrons and 
protons. Mendell obtained previously the general form of mutual friction 
damping from vortices, which incorporated the entrainment effect, 
in the case of type-II superconductivity \cite{MENDELL}.

If the proton superconductor is type-I, 
non-superconducting  domains coaxial with the neutron vortices 
nucleate in response to the entrainment current set-up by the
\begin{figure}[t] % fig 1
\begin{center}
\psfig{figure=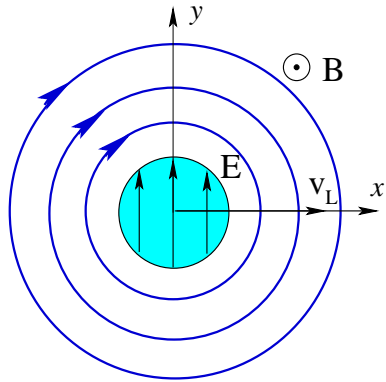,height=5.cm,width=5.cm,angle=0}
\end{center}
\caption{%(Color online) 
An illustration of the structure of a rotational vortex placed 
in a type-I superconductor. The vortex velocity field is shown 
by the concentric circles. The non-superconducting domain (shaded region)
of radius $a$ 
is coaxial with the  vortex and carries a magnetic 
field $H_{cm}\sim 10^{14}$ G. The vortex motion along the $x$-axis generates
a transverse electric field, which drives the electron current through the 
domain and leads to Ohmic dissipation.
}
\label{MSfig:fig1}
\end{figure}  
vortex circulation~\cite{SSZ}. Consider a cylindrical domain of radius
$a$ coaxial with a  vortex (see Fig. 1 for  an illustration). 
In the cylindrical coordinates ($r$, $\phi$, $z$)
with the symmetry axis at the center of the vortex the magnetic field 
induction is  [ref.~\cite{SSZ}, Eq. (20)] 
\be\label{eq:2} 
B_z(r) = 
\frac{\Phi_1}{2\pi\lambda^2}{\rm ln}\left(\frac{b}{a}\right)
\frac{N(r)}{N(a)}, \quad B_x = B_y = 0, 
\ee
where $\Phi_1 = (\rho_{pn}/\rho_{pp})\Phi_0$ ,
$\Phi_0$ is the flux quantum, $\lambda$ is the 
magnetic field penetration depth, $b$ is the vortex (outer) radius
and 
\be\label{eq:3}
N(r) = I_0\left(\frac{b}{\lambda}\right)K_0\left(\frac{r}{\lambda}\right)
-K_0\left(\frac{b}{\lambda}\right)I_0\left(\frac{r}{\lambda}\right),
\ee
where $I_0(z)$ and $K_0(z)$ are the modified Bessel functions. 
In the mean-field approximation the magnitude of 
the (non-quantized) flux $\Phi_1$  
is determined by the effective mass of a proton quasiparticle
$ \rho_{pn}/\rho_{pp}\equiv \vert m_p^*/m_p-1\vert$.

Now we can write down the proton supercurrent 
$j_{\phi} = (c/4\pi ) (\vec \nabla\times \vecB)_{\phi}$ by substituting 
the $B$-field from Eq. (\ref{eq:2}). We need, however,  the 
velocity of superfluid protons, which is the difference 
between the net supercurrent and the supercurrent
moving with the neutron superfluid velocity; we find
\be\label{eq:4}
v_{p \phi} = \frac{k\hbar}{2mr}\left[\frac{r}{\lambda}
{\rm ln}\left(\frac{b}{a}\right){\rm coth}
\left(\frac{b-r}{a}\right)-1\right].
\ee
Eq. (\ref{eq:4}) keeps the leading order term of the expansion of 
the Bessel functions with respect to large arguments $r/\lambda$, 
where $r$ is a mesoscopic scale $\ge a$.

Consider now a  vortex which moves at a constant velocity $\vecv_L$,
and carries a coaxial normal domain of protonic fluid with respect to the 
background electron liquid (see Fig. 1).
The equation of motion of the superfluid protons,
written in the reference frame where the  vortex is at rest,
acquires an additional term $(\vecv_L\cdot \bnabla )\cdot\vecv_p$.
The continuity of the electro-chemical potentials of the superfluid
and normal phases across the boundary of  the normal and
superconducting phases  gives $\mu_S = \mu_N$, where
\be
\mu_S = \mu-m_p^* v_L v_p(r) \frac{y}{r},\quad \mu_N = \mu -e\phi,
\ee
here $\mu$ is the proton chemical potential in equilibrium, $\phi$ is 
the scalar electric potential and we use  2d Cartesian 
coordinates with vortex circulation along $z$-axis and vortex velocity 
along the $x$-axis, see Fig. 1; small terms
$O(\Delta^2/\mu$) are neglected. Thus, the motion of the  vortex generates 
a constant transverse electric field across the normal domain 
\be\label{eq:6} 
E_y = -\nabla_y \phi = -\frac{m_p^*v_L v_p(a)}{ea}, \quad E_x=E_z=0.
\ee
The power dissipated per unit length of a vortex is  
$W = \sigma E^2\, (a/b)^2$, 
where $\sigma$ is the electrical conductivity and the factor 
$(a/b)^2$ is the fractional 2d volume occupied of the domain.
Upon substituting Eq. (\ref{eq:6}) in this relation, 
we obtain an alternative form of dissipation
$W = \eta v_L^2$, which identifies the friction coefficient
\be\label{eq:7}  
\eta = \frac{\sigma}{c^2}
\left(\frac{\Phi_1}{2\pi ab}\right)^2
\left[\frac{a}{\lambda}{\rm ln}\left(\frac{b}{a}\right)
{\rm coth}\left(\frac{b-a}{a}\right)-1
\right]^2.
\ee
Equation (\ref{eq:7}) is our central result, which defines the friction 
coefficient for a vortex featuring a single coaxial domain of 
a type-I proton superconductor in terms of the electrical conductivity 
of the electron Fermi-liquid in normal proton matter. 

Since the mean-free path of electrons is much smaller than the size of 
a single normal domain, we can neglect the finite-size effects and 
use the result for the bulk normal matter \cite{BBP}. 
The zero-field conductivity of ultra-relativistic
electrons is $\sigma_0 = n_e e^2 c\tau_c/(\hbar k_F)$,
where the relaxation time for the Coulomb scattering of electrons off 
the protons in the normal domains is \cite{BBP} 
\be\label{eq:8} 
\tau_c = \frac{12}{\pi^2}
\left(\frac{\hbar c}{e^2}\right)^2 
\left(\frac{\epsilon_F}{T}\right)^2 \frac{k_{FT}}{ck_F^2},
\ee
where $\epsilon_F$ and  $k_F$ are the electron Fermi-energy
and Fermi-wavenumber, and $k_{TF} = [4k_Fm_p^*e^2/\pi\hbar^2]^{1/2} $
is the Thomas-Fermi wave-length. Since the Larmor radius of an electron
moving in a normal domain carrying a field $H_{\rm cm} \sim 10^{14}$ G
is much smaller than the linear size of the domain, the 
conductivity becomes
%\be\label{eq:9} 
$\sigma = {\sigma_0}/{(\omega_c\tau_c)^2}, $ where $\omega_c 
= ({eH_{cm}})/({\hbar k_F}),$
%\ee
where $\omega_c$ is the electron cyclotron frequency.
Table I lists some of the relevant parameters of the problem, 
including transport quantities and the drag-to-lift ratio 
at the temperature  $T = 10^8$ K. 
The kinetic coefficients can be rescaled to other temperatures by using the 
scalings  $\tau \propto T^{-2}$, $\sigma \propto T^{2}$ and
$\zeta \propto T^{2}$. The drag-to-lift ratio satisfies the condition 
$\zeta \ll 1$ at all temperatures
(the largest values correspond to temperatures just below the  
critical temperature $T_c \sim 10^9$ K).
We shall return to the implications of these results for 
the neutron star precession in the closing section.
\begin{table}
\caption{Listed are the protons density (column 1), the Fermi-wave
number (2), the effective mass of protons (3), 
the magnetic field penetration depth (4), the critical
thermodynamic field (5), the electron relaxation time (6), the electrical 
conductivity (7), and the drag-to-lift ratio (8).
The short-hand notions $\rho_{14} = \rho/10^{14}$, etc., are used.}
\begin{tabular}{cccccccc}
\hline
$\rho_{14}$ & $k_{Fp}$ & $m_p^*/m_p$ & $\lambda$ & $H_{cm,\, 14}$ 
& $\tau_{c,-13}$ & $\sigma_{16}$ & $\zeta_{-4}$\\
g cm$^{-3}$    & fm$^{-1}$ &  & fm & G & s & s$^{-1}$ & \\
\hline
7.91 & 0.85 & 0.69 & 41.58 & 11.62 & 10.94 & 3.76 & 0.39\\
8.31 & 0.88 & 0.68 & 39.20 & 7.82 & 12.20 & 8.56 & 0.36\\
8.56 & 0.90 & 0.68 & 37.87 & 3.88 & 12.90 & 35.94 & 0.88\\
\hline
\end{tabular}
\end{table} 

\section{Dynamics of 1+N structures}
\label{1+N}
Buckley et al \cite{BUCKLEY} argued (qualitatively) that the size of the 
normal domains could be large enough to accommodate about 
$N = 10$ neutron vortices across a single normal domain of protonic fluid. 
Since there is  no dynamical coupling (in the sense of the entrainment) 
between the vortices and the normal domains 
the damping of the differential rotation between electron-proton plasma
and the neutron superfluid is due to the interaction of domain 
(non-superconducting) protons with  the core quasiparticles 
confined in the neutron vortex core. The relaxation process 
is thus the same as for the case where the proton fluid is 
non-superconducting over the entire bulk of the core,  but the final 
result needs to be rescaled by the ratio of the areas occupied by 
the normal and superconducting layers. The relaxation time 
per single vortex is \cite{SEDRAKIAN98}
\be\label{eq:10} 
\tau_{np} = 6
\left(\frac{k_{Fp}}{k_{Fn}}\right)^4\frac{m_n\mu_{pn}^*}
{\hbar m_p^*T\sigma_{np}}
~{\rm exp}\left(0.02 \frac{\Delta_n^2}{\epsilon_{Fn}T}\right),
\ee
where $k_{Fp}$ and $k_{Fn}$ are the Fermi-wave-numbers of protons and 
neutrons, $\mu_{pn}^* = m_p^* m_n^*/(m_p^* +m_n^*)$ is the reduced 
effective mass, with $m^*_n$ begin the neutron effective mass,  
$\sigma$ is the total in-medium neutron-proton scattering
cross-section, $\Delta_n$ is the gap in the neutron quasiparticle spectrum, 
$\epsilon_{Fn}$ is the neutron Fermi-energy. [Eq. (\ref{eq:10})
differs from the analogous expression in ref. \cite{SEDRAKIAN98} by
the factor $4m_n/\hbar P$; here $P$ is the pulsar period,
$m_n$ - free-space neutron mass].

In the relaxation time-approximation, 
the force exerted by the normal proton on a single vortex 
is given by a phase-space integral 
\be\label{eq:12} 
\vecf = \frac{1}{\tau_{np}}\int 
\frac{d\vecp}{(2\pi\hbar)^3}~ f_F(\vecp, \vecu) = \eta \vecu,
\ee
over the proton Fermi-distribution function, 
$f_F(\vecp, \vecu) = 
\left[{\rm exp}(\epsilon_p-\epsilon_{Fp}+\vecp\cdot \vecu)/T
+1\right]^{-1},
$
where the quasiparticle energy shifted due to the motion with a
velocity $\vecu$. Here $\epsilon_p$ is the dispersion relation of normal
protons, $\epsilon_{Fp}$ is their Fermi-energy.
The integral (\ref{eq:12}) is straightforward in the $T=0$ 
limit and we obtain
\be
\eta = \frac{\hbar k_{Fp} n_p}{c\tau_{np}},
\ee
where $n_p$ is the proton number density.
\begin{table}
\caption{The columns 3-5 list the relaxation time for proton scattering 
off the neutron quasiparticles in the neutron vortex cores (3), the 
vortex friction coefficient (4), and the drag-to-lift ratio (5) for 
the temperature $T = 10^8$ K. 
The columns 7-9 list the same parameters for $T = 10^7$ K. 
}
\begin{tabular}{cccccccc}
\hline
$\rho_{14}$ 
& $k_{Fp}$ 
& $\tau_{np}$ 
& $\eta_{10}$ 
& $\zeta_{-2}$
& $\tau_{np}$ 
& $\eta_{10}$ 
& $\zeta_{-2}$\\
g/cm$^{3}$    
& fm$^{-1}$ 
& s 
& g/cm/s
& 
& s
& g/cm/s
&
\\
\hline
&  
& $T_8 =1$ 
&  
&  
& $T_8 =0.1$ 
& 
& \\
\hline 
0.40 & 0.89 & 71.48 & 10.48 & 7.95 & 285.9 & 2.62 & 19.87\\
0.60 & 1.08 & 428.9 & 3.68 & 1.86 & 1072.2 & 1.47 & 7.45\\
0.80 & 1.26 & 786.3 & 3.76 & 1.42 & 1572.6 & 1.88 & 7.13\\
\hline
\end{tabular}
\end{table}
The result for the friction coefficient $\eta$ and the corresponding
drag-to-lift ratio for several densities are listed in Table II for 
the case where the proton fluid is non-superconducting. For a given 
model of the type-I superconducting 
structure, the friction coefficient $\eta$ must be rescaled by 
a factor $(d_N/d_S)^2$. 

\section{Implications for Precession and Conclusions}
\label{conclude}

The SWC no-go theorem requires the condition $(I_S/I_N)\zeta < 1$ 
to be fulfilled for precession to occur; otherwise the precession 
is damped. The magnitude of the ratio $I_S/I_N$ depends on the 
superfluid-normal fluid friction within  all superfluid
regions of a neutron star and is difficult to access. Glitches and 
post-glitch relaxation provide a model independent lower bound on 
$I_S/I_N \ge 0.1$. An upper bound is difficult to place, since the 
deep interiors of neutrons stars, if superfluid, 
could be decoupled from the observable parts of the star on 
evolutionary timescales without any effect on short time-scale physics 
\cite{SED_CORDES}
(but one needs $\zeta \to 0$, rather than $\zeta\to\infty$,
to prevent the damping of the precession). However, it is 
rather unlikely that this ratio exceeds unity by many orders of 
magnitude. For the first dissipation channel studied
$\zeta \sim 10^{-4}-10^{-3}$ (Table I) and this clearly 
suggests an undamped precession.
The second chanel is more effective, $\zeta \sim 10^{-2}$
(Table II), but these numbers must be reduce by a  
factor $(d_S/d_N)^2 \simeq 100$ . On account of the lower bound on the 
ratio of the moments of inertia, one can conclude that the precession 
is undamped for both dissipation mechanism.

We have provided a first discussion of the dynamics of the type-I
superconducting domains in neutron star interiors. Although the 
details of the coupling of the superconducting and normal components 
of the star depend on  the form of the superconducting-normal
structures that nucleate and, in particular, whether the 
non-superconducting domains are dynamically coupled 
to the rotational vortices or not, in all cases we find friction coefficients
that imply undamped precession. 

The results above by no means suggest that 
type-I superconductivity is the only resolution to the 
precession puzzle and alternatives should be searched for.
An alternative to free, Eulerian precession is the 
forced precession due to time-dependent periodic torques \cite{SWC}. 
Other periodic motions, for example, Tkachenko oscillations of the
vortex lattice could generate the observed 
timing features \cite{TKACH,RUDERMAN,BAYM}. While the eigen-frequencies
of the Tkachenko modes are of correct order of magnitude and 
could explain long-term periodicities it remains to be studied whether these 
modes will be undamped by the mutual friction between the superfluid 
and the normal fluid.

The fact that statistically  insignificant number 
of pulsars show long-term variabilities, indicates that a 
subtle tuning is needed for the
underlying mechanisms to work. On the other hand,
the Eulerian precession, if undamped, should be a common place
in the pulsar population, since  neutron stars frequently undergo 
non-axisymmetric perturbations such as glitches and quakes.

\section*{Acknowledgments}
I am grateful 
to A. R. Zhitnitsky for discussions and Ira  Wasserman for reading 
the draft and comments. I would like to thank the Institute for Nuclear 
Theory at the University of Washington for its hospitality and the 
Department of Energy for the partial support of the stay at the UW.

\end{document}